# IoT Malware Network Traffic Classification using Visual Representation and Deep Learning

Gueltoum Bendiab*, Stavros Shiaeles*, Abdulrahman Alruban[†], Nicholas Kolokotronis[‡]

*Cyber Security Research Group, University of Portsmouth, PO1 2UP, Portsmouth, UK
gueltoum.bendiab@port.ac.uk, sshiaeles@ieee.org

[†]CSCAN, University of Plymouth, PL4 8AA, Plymouth, UK
abdulrahman.alruban@plymouth.ac.uk

[‡]Department of Informatics and Telecommunications, University of Peloponnese
22131 Tripolis, Greece. nkolok@uop.gr

*Abstract*—With the increase of IoT devices and technologies coming into service, Malware has risen as a challenging threat with increased infection rates and levels of sophistication. With- out strong security mechanisms, a huge amount of sensitive data is exposed to vulnerabilities, and therefore, easily abused by cybercriminals to perform several illegal activities. Thus, advanced network security mechanisms that are able of performing a real-time traffic analysis and mitigation of malicious traffic are required. To address this challenge, we are proposing a novel IoT malware traffic analysis approach using deep learning and visual representation for faster detection and classification of new malware (zero-day malware). The detection of malicious network traffic in the proposed approach works at the package level, significantly reducing the time of detection with promising results due to the deep learning technologies used. To evaluate our proposed method performance, a dataset is constructed which consists of 1000 pcap files of normal and malware traffic that are collected from different network traffic sources. The experimental results of Residual Neural Network (ResNet50) are very promising, providing a 94.50% accuracy rate for detection of malware traffic.

## 1. INTRODUCTION

During the last years, there is a tremendous increase of Internet of Things (IoT) devices usage in several fields, ranging from industry, health, automation and education to smart homes and smart cities [1]. The current predictions on the number of connected IoT devices are expected to surpass 50 billion smart objects by 2020 [2]. This number is projected to reach 75.44 billion worldwide by 2025 [3]. Further studies confirm that the network of connected "smart" devices is the next major step in delivering Internet's promise of making the world a connected place [1], [3]. However, this technology also comes with new security and privacy challenges and it is not as safe as it looks. In fact, IoT networks usually rely on low- cost devices (e.g. temperature sensors, surveillance cameras, etc.) with limited resources (i.e. low power sources, limited amount of memory and processing power), and therefore, weak or no security embedded into them [1], [4]. These limitations hinder the execution of complexes security tasks on those devices and give malicious actors the opportunity to easily compromise them and perform several illegal activities, a situation which abuses the security and integrity of the devices and the network. In this context, a recent study by the cybersecurity firm Avast found that two out of five IoT devices are exposed to cyberattacks [5]. The report confirms that Botnets are the most common type of attacks when an IoT device is compromised [5], either standalone or aggregated to become part of a botnet, capable of launching destructive DDOS (Denial of service) attacks. The Mirai botnet [6] is an example of IoT powered DDoS attacks that took advantage of insecure IoT devices to conduct massive security breaches. This attack took the Internet by storm in late 2016 [7] and took down hundreds of services such as DNS-providers, GitHub, Amazon, Twitter, Netflix, Reddit, etc. [6], [7]. Unlike known botnets, which are typically made of computers, the Mirai botnet is largely made of IoT devices such as digital cameras, DVR players, temperature sensors, etc. [7], [8]. Investigations estimated that more than 400,000 infected IoT devices were involved in this massive DDoS attack, which makes it the most powerful DDoS attacks in history.

Most techniques used for detecting such malware network traffic rely on databases of known attack signatures [9], where the incoming traffic is compared against the predefined malware signatures to identify possible attacks [9], [10]. Those techniques are highly accurate and very effective at detecting known attacks, but largely ineffective in detecting unknown and new versions of the emerging threats, for which there exist no signatures [4], [10]. Furthermore, they require considerable resources and overhead, and manual interventions to updates the attack signatures [4], [9]. Therefore, they are not suitable for real-time network anomaly detection. Great effort has been directed to overcome these limitations and a variety of approaches that focuses on behaviour analysis or anomaly-detection has been proposed [9]. However, until now there is no approach or system, which perfectly can detect or dynamically adapted to distinguish between legitimate and malware traffic, especially unknown and new malware. There- fore, in this research work, we propose a novel approach to classify network traffic by using visual representation and the deep learning algorithm, Residual Neural Network (ResNet) [11]. In this approach, the incoming traffic is collected and converted to 2D images by using the visual representation tool Binvis [12]. Then, by leveraging the Residual Neural Network algorithm, the system is trained to identify and distinguish potential malware and legitimate network traffic. This enables an automated and faster process of detecting known and unknown attacks.

This work is an extension of our previous works in [13] and [14] by using Residual Neural Network with more samples of malware and legitimate pcap files for the training and testing phases. Work in [13], centred on detecting malware executables as opposed to traffic by using Self-Organizing Incremental Neural Networks (SOINN), while in [14], we used the convolutional neural network MobileNet for IoT malware traffic analysis. Furthermore, the two approaches were tested on small datasets (only 100 samples), which leads to restricting neural network training options. In this work, the efficiency of the three classifiers is tested on a dataset that is consists of 1000 Binvis images of legitimate and malware pcap files that are collected from different network traffic sources. According to the experimental and comparative results with the classification models in [13] and [14], Residual Neural Network (ResNet50) algorithm gives the best performance with 94.50% overall accuracy rate for detection of malware traffic.

The overall structure of the paper is organised as follows: Section 2 summarizes the prior works done in malware traffic analysis and classification using the machine-learning technique. In Section 3, we give an overview of the proposed approach and the dataset used in the experiments to prove its effectiveness. Section 4 presents experiment results and analysis as well as a comparison with other methods. Finally, Section 5 concludes the paper and present future work.

## 2. RELATED WORK

Techniques used for detecting malware network traffic are generally divided into two main categories: signature-based detection and behaviour or anomaly-based detection [4], [10], [15]. Signature-based detection techniques have been used since the earliest days of network security monitoring. They refer to databases of known attack signatures, where for each specific threat, a pattern (or a signature) that identifies its unique characteristics is created, so that specific threat can be identified in the future [9], [10]. Then, the signatures are compared against incoming traffic to identify possible attacks [9]. Signature-based techniques are good at detecting known attacks, but largely ineffective in detecting unknown and new versions of attacks for which do not exist signatures. In fact, signatures can only identify threats that are already known [4], [10] and a frequent update of attack signatures needs to be performed [4] in order to be current. However, this might require considerable resources and overhead, and manual interventions [4], [9].

In the face of signature-based techniques limitations, researchers are now focusing on behaviour analysis or anomaly- detection techniques [9], [16]. In this context, many re- searchers have argued for the importance of machine learning in malware traffic classification and intrusion detection, and several network-level anomaly detection solutions have adopted the machine learning approach [10], [17]. These approaches analyse the available information of the network traffic, by extracting features that can be used to distinguish the malware traffic from the legitimate one [16], [17]. Then, they use these features to train the classification model to detect potential attacks [4], [17]. The output results are generally presented in a binary fashion (i.e. normal or malware), therefore, it will label each data instance as either normal or anomaly. For instance, study in [18] compared the predictive accuracy of five supervised machine-learning algorithms with five features selection sets derived from four previous work done in android malware traffic detection. The five-machine learning classifier are Naïve Bayes (NB), K-nearest Neighbour (KNN), Decision Tree (J48), Multi-Layer Perceptron (MLP) and Random Forest (RF). The experimental results showed that Multilayer perceptron (MLP) classifier using the features selection set derives from the features selection method, out- performs all other classifiers with 83% accuracy rate, 90% True Positive (TP) rate and 23% False Positive (FP) rate. In [19], the authors investigated the ability of Recurrent Neural Networks (RNN) to detect the behaviour of network traffic by modelling it as a sequence of states that change over time. The network traffic features were transformed into a sequence of characters and then RNNs is used to learn their temporal features. From the experimental results, authors concluded that the RNN detection models have problems for dealing with traffic behaviours not easily differentiable as well as some special cases of imbalanced network traffic. Another recent work in [20], applied seven different machine-learning algorithms with the well-known dataset "Kyoto 2006+" that contains 24 features [21]. The seven learning algorithms applied are K-Means, K-Nearest Neighbours (KNN), Fuzzy C- Means (FCM), Support Vector Machine (SVM), NaiveBayes (NB), Radial Basis Function (RBF) and Ensemble method that combines the above-mentioned six algorithms. Experimental results showed that most of the learning algorithms provided a satisfying accuracy of over 90%.

In the same context, authors in [22] proposed a self-learning anomaly detection approach, which can be adapted to changes in the network traffic. They used the Discriminative Restricted Boltzmann Machine (DRBM) neural networks, which was trained only on normal traffic, and the knowledge about anomalous traffic evolved dynamically to recognise abnormal traffic by the DRBM classifier with high accuracy. Two sets of experiments were conducted to demonstrate the effectiveness of this approach. In the first one, they collected real traffic traces from a normal network host and an infected network host. While, in the second, they used the public dataset KDD'99 [23]. Authors stated that the classifier obtained the best accuracy results with the first set of experiments (92%-96%). In another pertinent work [24], authors focused on detecting anomalies in conflicting network environment by filtering out noisy data associated to irrelevant features. Thus, they used a combination of machine learning algorithms for features selection. Firstly, they used the unsupervised technique k-means clustering to identify clustered features. Then, they used the Naive Bayes algorithm and the Kruscal- Wallistest [25] for features ranking. In this step, only relevant features were chosen according to their ranking. In the last step, they used the C4.5 decision tree classifier to evaluate the selected features in the second step. From the experimental results, authors found that reducing the number of features helped in decreasing the amount of computation needed and therefore, enhancing the anomaly detection speed and accuracy.

In more recent study [26], authors introduced a transfer- learning model for network traffic classification, where the maximum entropy (Maxent) method [27] was applied as the base classifier for the model. Unlike traditional machine leaning techniques, Maxent is used to transfer the knowledge from source domain into target domain in traffic classification and therefore, preserve the performance of traffic classification when the network environment change. The efficiency of this model was tested on two different traffic datasets (training and testing) that were collected at the University of Cambridge, and it obtained an average accuracy over 98% with the testing dataset. In the same context, authors in [28] proposed a semi- supervised method to detect malware network traffic by utilising density models, which are based on recent advances in deep generative models and variational inference theory [29]. Representation features of the raw flows were automatically extracted in an unsupervised way by using the variational auto-encoder (VAE). Then, related flows were clustered into a latent feature space, which makes the classification more accurate. This approach used only a few labelled flows but achieved a satisfying accuracy of over 90%.

## 3. PROPOSED APPROACH

### A. Approach Overview

As illustrated in Fig. 1, the proposed approach for IoT malware traffic analysis consists of two main steps, first obtaining the corresponding visual representation of the collected network traffic, and second, processing this visual representation by the trained classification model. The network traffic collection is done by using pcap files containing pre- captured network traffic (i.e. normal and abnormal traffic) that can be replayed through TCPreplpay in order the sniffer to capture packets to files. Then, for performance reasons, multiple packets chunks are created and forward to the visual representation tool to convert them into a 2D image as the first step of the procedure. In the second step, the Residual Neural Network (ResNet50) [11] is used to analyse the produced images against its in-depth training. ResNet-50 is a pre-trained deep learning model for image classification. It used a Convolutional Neural Network (CNN, or ConvNet), which is a class of deep neural networks, most commonly applied to analysing visual imagery. ResNet-50 is 50 layers deep and is trained on a million images of 1000 categories from the ImageNet database. Furthermore, the model has over 23 million trainable parameters, which indicates a deep architecture that makes it better for image recognition. ResNet has the advantage to allow training deep neural networks with hundreds or even thousands of layers (more than 150 layers) and still achieves compelling performance [11]. Because of this powerful representational ability, ResNet quickly became one of the most popular architectures in various computer vision tasks such as image classification, object detection and face recognition [11], [30].

In our approach, the topological structure of the neural network is built in the training step, while in the testing step; the collected traffic is tested against the samples in the database to perform the classification. Detected malware traffic will be used to continuously train the classifier in order to enhance its detection accuracy.

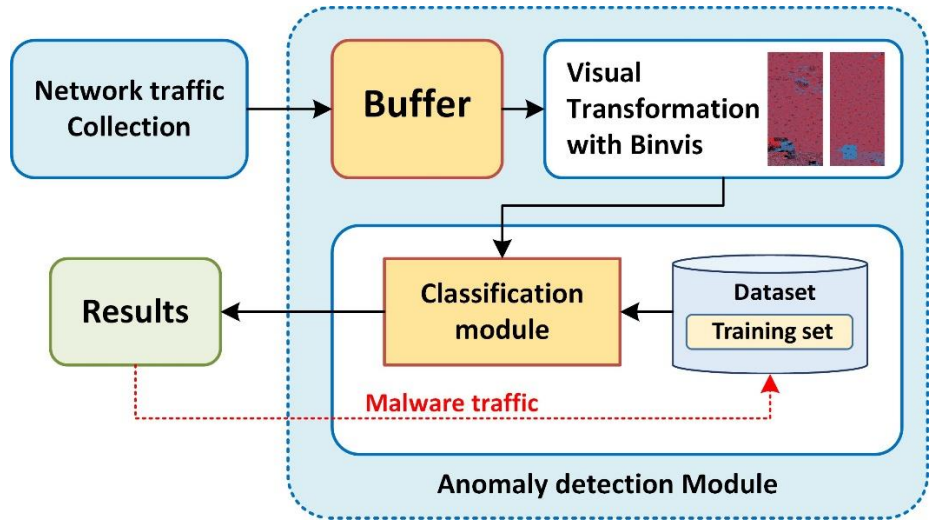

Fig. 1. High-level architecture of the proposed approach.

## B. Visual transformation of the network traffic

The collected traffic is stored and converted into RGB images by using the visual representation tool Binvis [12]. This tool is widely used, especially by security researches, to visualize binary-file structures in unique ways, in order to detect and analyse malicious content [12], [31]. In the conversion process, the pcap chunk files are stored in binary format and then mapped to appropriate RGB values. The RGB values mapping is done by comparing each byte value in the pcap file to their equivalent in the ASCII table, according to the predefined colour scheme. Binvis divided the different ASCII bytes into four groups of colours, where printable ASCII bytes are assigned a blue colour, control bytes are assigned a green colour and extended ASCII bytes are assigned a red colour. Black (0x00) and white (0xFF) colour respectively represent null and (non-breaking) spaces [31]. After that, the coordinates of each byte colour in the output RGB image are identified using a clustering algorithm based on space-filling curves [32] to ensure grouping closer data together. This clustering algorithm outperforms other curves in preserving the locality between objects in multi-dimensional spaces, which helps to create much more appropriate RGB images for the classification process. The size of the output RGB image is 784 (1024*256) bytes.

Fig. 2 shows Binvis images for normal and malware pcap files, which are created using the Hilbert space-filling curve [32]. Positive results can be concluded from these images, as differences between a legitimate and malware pcap file are visible. As seen in the figures, images of malware pcap files have more predominance of black (Null Bytes) or white areas (Spaces). Whereas normal traffic can be recognised by the distribution of ASCII characters or colours across the image.

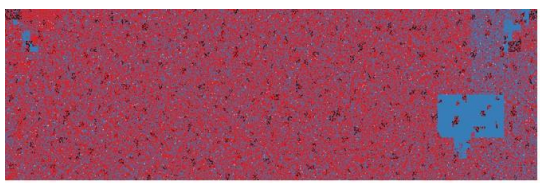

Binary visualisation of normal pcap file

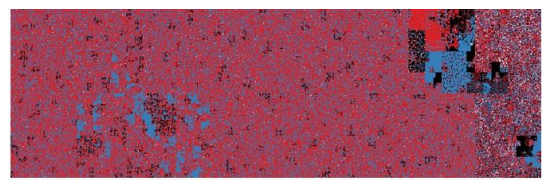

Binary visualisation of botnet pcap file

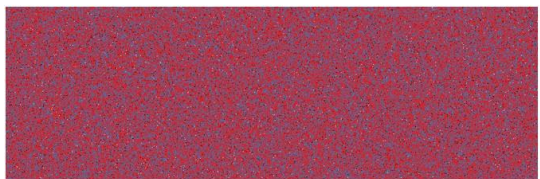

Binary visualisation of normal pcap file

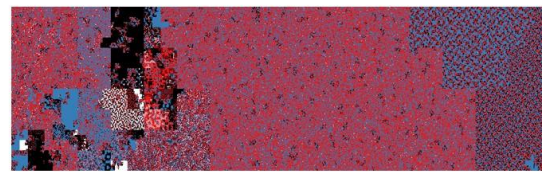

Binary visualisation of backdoor pcap file

Fig. 2. Binvis images of normal and malware pcap files created with the Hilbert space-filling curve.

## C. Data Collection

By using the proposed visual transformation approach, we created our own dataset for training and testing the neural network. The dataset includes a mixture of 1000 Binvis images of normal and malware traffic that were collected from different network traffic sources. Normal PCAP files contain captured normal traffic from the Cyber-trust project network and other sources. The traffic was collected from various clean devices in the network using tools such as Nmap and Wireshark. Malicious pcap files were collected from three main public sources of malware PCAP files including the malware traffic analysis repository[1], the NETRESEC repository[2] and the malware datasets of the stratosphere lab[3]. The malware pcap files contain real malicious traffic that was generated by different types of attacks such as trojans, botnets, IoT based attacks (DDoS, Key loggers, OS scans, spyware), backdoors, etc. Fig. 3 shows the percentage of malicious traffic samples of the whole dataset.

The ResNet algorithm was trained by 800 different BinVis images (normal or malware) with a size of 784 (1024*256) bytes. The images are labelled as normal or malware. Other 200 different BinVis images were used for testing the classifier. The testing set represents the unknown network traffic that we want to classify.

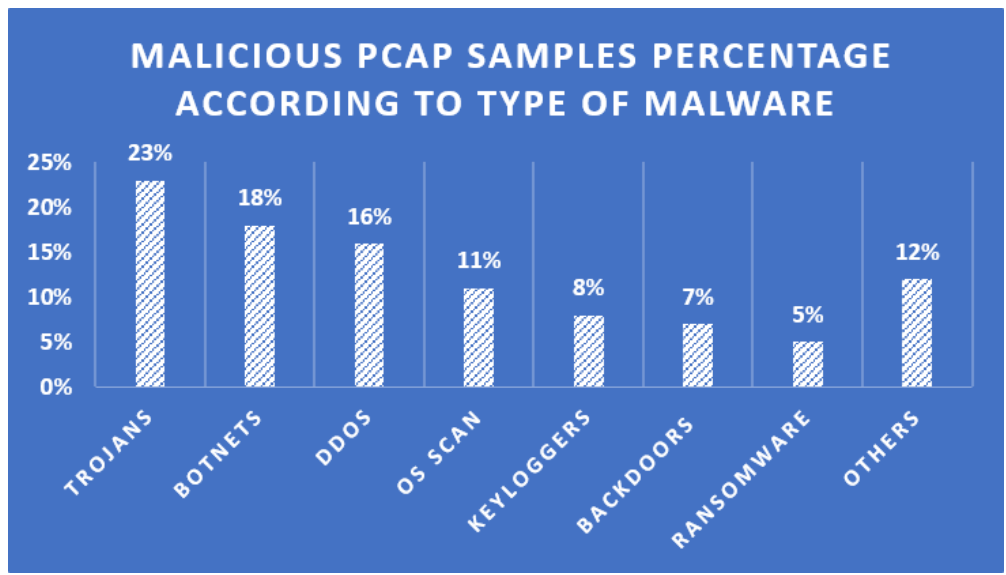

**Fig. 3. Malicious traffic samples percentage according to type of malware.**

## 4. EXPERIMENTAL RESULTS

This section presents the results of the experiments con- ducted over the proposed approach in order to demonstrate its effectiveness and reliability. The performance metrics used to evaluate the performance of the classifier are accuracy (*A*), precision (*P*), recall (*R*), and F-score (*F1*). The accuracy refers to the ratio of all correctly classified samples either normal or malware. Precision (P) provides the percentage of positively classified samples that are truly positive. Recall represent the number of normal samples that were correctly classified, while F-score is a weighted average between precision and recall. Table I illustrates the equation used for each metric. True Positive (*TP*) is the number of instances correctly classified as malware traffic, True negative (*TN*) is the number of instances correctly classified as normal traffic, False Positive (*FP*) is the number of instances incorrectly classified as malware traffic, and False Negative (*FN*) is the number of instances incorrectly classified as normal traffic.

---

[1] https://github.com/tatsui-geek/malware-traffic-analysis.net
[2] https://www.netresec.com/?page=PcapFiles
[3] https://www.stratosphereips.org/datasets-malware

**Table 1: Metrics used to evaluate the performance of the proposed solution.**

| Metric | Equation |
|---|---|
| Accuracy (A) | $A = \frac{TP+TN}{TP+TN+FP+FN}$ |
| Precision (p) | $P = \frac{TP}{TP+FP}$ |
| Recall (R) | $R = \frac{TP}{TP+FN}$ |
| F-score (F1) | $F1 = 2 \times \frac{P \times R}{P+R}$ |

The experiments were conducted on a physical machine, running on Intel Core i7 CPU, 3.80 GHz, with 32 GB memory and the Windows 10 enterprise 64 bites OS. An NVIDIA GTX 1060 GPU with 6 GB memory is used as an accelerator. The Resnet50 learning algorithm was implemented using the open-source Fastai Python library[4] that was developed at the University of San Francisco for deep learning. Based on top of PyTorch, Fastai contains some of the most popular algorithms for image classification and natural language tasks.

### A. Test Results

Several tests were carried out to evaluate the success of the detection method and determine the accuracy of the proposed classifier. Fig. 4 and Fig. 5 illustrate the results of the Resnet50 Neural Network training and testing in terms of accuracy, training, and validation loss. The errors loss (i.e. training and validation loss) is a summation of the errors made for each epoch in training or validation (i.e. testing) sets and it is expected to decrease after each or several epochs. The training of the NN should be stopped if, the validation loss should be similar to or slightly higher than the training loss. However, if the validation loss is lower than the training loss one, we should keep doing more training (i.e. under-fitting). In our experiments, the training of the neural network is done for 50 epochs with a batch size of six. The learning rate parameter is fixed to 0.05 after running the learning rate finder function (LRFinder). This function eliminates the need to perform numerous experiments to find the optimal values [33]. As can be seen in Fig. 4, after 50 epochs, Resnet50 achieved an overall accuracy of 94.50% with very close validation loss (0.237579) and training loss (0.23048). In all the previous epochs, the validation loss was lower than the training loss with a difference ranging between 0.11 and 0.03, while the accuracy values were between 91.22% and 95.32%.

Fig. 5 presents the overall results of the proposed approach, which reached an overall detection accuracy of 94.50%, which is a high rate and meets the required accuracy rate in practical use. The precision of classifier is also very high with a rate of 95.78%, which shows strong overall confidence in the pattern recognition process. The recall rate was lower than the precision rate (94.02%), which shows the efficiency of Resnet50 in the correct classification of most of the samples. Knowing that in our work, precision is more important than recall because getting False Negatives (FN), when malware traffic is considered as normal, cost more than False Positives (FP), when normal traffic is considered as malicious traffic. Based on these results, the F-score value (F1) is 94.90%.

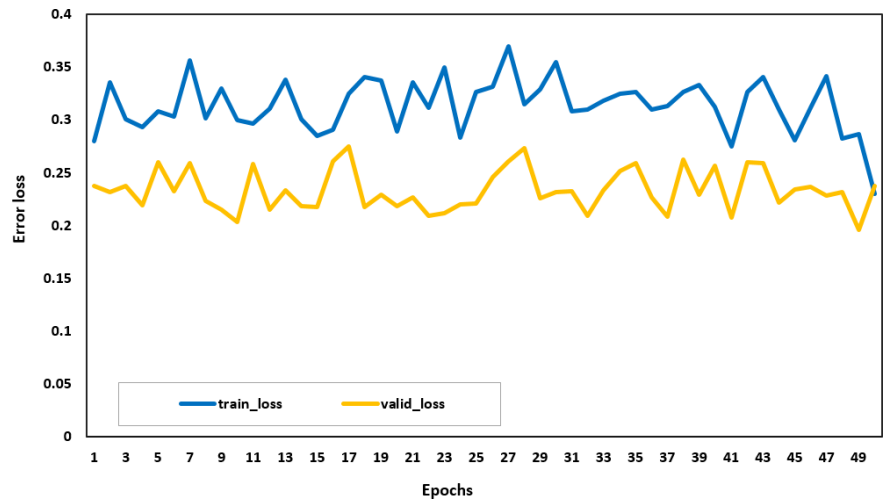

**Fig. 4. Error loss values during the training and testing of Resnet50**

---

[4] https://www.fast.ai/2018/10/02/fastai-ai/

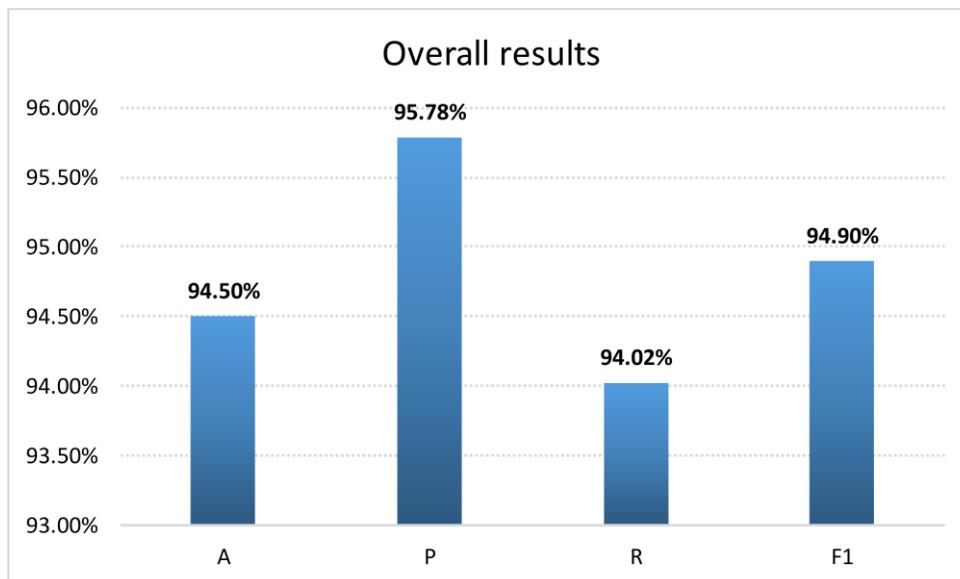

**Fig. 5. Overall results for Resnet50.**

### B. Comparison

In order to prove the effectiveness of Resnet in detecting malware traffic, a comparison with the learning algorithms used in our previous work [13], [14] is done based on the metrics defined in this section. In [13], we used self-organizing incremental neural networks (SOINN) for the analysis and detection of malicious payloads. SOINN is an unsupervised machine learning algorithm known for its incremental abilities, provides the ability to learn fast by exploiting only the information needed for building the neural network (NN) and removing redundant nodes [13]. In [14], the same approach is used with the convolutional neural networks MobileNet proposed by Google.

After the optimum initial configuration was found for each classifier, the learning algorithm was retrained on the created training dataset and assessed against the testing dataset. In these experiments, the Residual CNNs ResNet was tested for 34-layers (ResNet 34) and 50-layers (ResNet 50). The same datasets were used for training and testing all the algorithms. The evaluation results are reported in Table II. From the obtained results, the Resnet50 has the best overall performance compared with other algorithms, with higher accuracy (94.50%) and precision (95.78%). It is also observed that Resnet performs better with more layers (Resnet50).

**Table 2: Comparison with other learning algorithms**

| Learning Algorithms | Accuracy | Precision | Recall | F-score |
|---|---|---|---|---|
| Resnet34 | 92.39% | 93.57% | 64.55% | 76.40% |
| Resnet50 | 94.50% | 95.78% | 94.02% | 94.90% |
| MobileNet | 91.32% | 91.67% | 91.03% | 91.35% |
| SOINN | 91.75% | 89.68% | 95.52% | 92.50% |

## 5. CONCLUSION

In this paper, we introduced a novel approach that utilises machine learning and visual representation to identify malicious network traffic. Form the experiments and comparison with multiple neural networks, the Residual Neural Network with fifty layers (ResNet50) has proved that it is the most effective in the identification of malware network traffic with an overall accuracy of 94.50%.

In the future, we plan to improve this work by using more samples to properly train and test the neural network, which undoubtedly will improve the predictive accuracy of the classifier. Furthermore, we intend to implement the machine- learning module in the intrusion detection systems (IDS) Suricata or snort in order to improve the protection and mitigation processes in this IDS. Another future work is studying the performance of the proposed approach in a real environment (e.g. time to obtain visual representation and classification), especially for online training.

## ACKNOWLEDGMENT

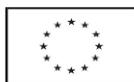
This project has received funding from the Euro- pean Union's Horizon 2020 research and innovation programme under grant agreement no. 786698. The work reflects only the authors' view and the Agency is not responsible for any use that may be made of the information it contains.